\documentclass[useAMS,usenatbib]{mn2e}

\usepackage{graphics}

\title[Upper limit on the mass of the central black hole in NGC 6388]{Radio observations of NGC 6388: an upper limit on the mass of its central black hole}
\author[D. Cseh, P. Kaaret, S. Corbel, E. K\"{o}rding, M. Coriat, A. Tzioumis, B. Lanzoni]{D. Cseh,$^{1}$\thanks{E-mail:
david.cseh@cea.fr} P. Kaaret,$^{2}$ S. Corbel,$^{1}$ E. K\"{o}rding,$^{1}$ M. Coriat,$^{1}$ A. Tzioumis,$^{3}$
 \newauthor B. Lanzoni$^{4}$\\
$^{1}$Laboratoire Astrophysique des Interactions Multi-echelles (UMR 7158),\\
\,\,\,CEA/DSM-CNRS-Universite Paris Diderot, CEA Saclay, F-91191 Gif sur Yvette, France\\
$^{2}$Department of Physics and Astronomy, University of Iowa, Van Allen Hall, Iowa City, IA 52242, US\\
$^{3}$Australia Telescope National Facility, CSIRO, PO Box 76, Epping NSW 1710, Australia\\
$^{4}$Dipartimento di Astronomia, Universita` degli Studi di Bologna, via Ranzani 1, I- 40127 Bologna, Italy\\}
\begin{document}

\date{Accepted 2010 March 22.  Received 2010 March 22; in original form 2010 February 16}

\pagerange{\pageref{firstpage}--\pageref{lastpage}} \pubyear{2010}

\maketitle

\label{firstpage}

\begin{abstract}
We present the results of deep  radio  observations with the Australia Telescope Compact Array (\textit{ATCA}) of the globular cluster NGC 6388.  We show that there is no radio source detected (with a r.m.s. noise level of 27 $\mu$Jy) at the cluster centre of gravity or at the locations of the any of the \textit{Chandra} X-ray sources in the cluster. Based on the fundamental plane of accreting black holes which is a relationship between X-ray luminosity, radio luminosity  and black hole mass, we place an upper limit of $\sim$1500~M$_{\odot}$ on the mass of the putative intermediate-mass black hole located at the centre of NGC 6388.  We discuss the uncertainties of  this upper limit and the previously suggested black hole mass of 5700~M$_{\odot}$ based on surface density profile analysis. 

\end{abstract}


\section{Introduction}

Following the early  discoveries of X-ray sources in globular clusters in the mid-1970s (Clark 1975; Clark, Markert \& Li 1975), it was proposed that the X-ray emission of these clusters was due to accretion of intracluster material released by stellar mass loss onto central black holes (Bahcall \& Ostriker 1975; Silk \& Arons 1975).  This started a debate about whether globular clusters contain black holes of intermediate masses (i.e. greater than the $\sim$ 30 M$_{\odot}$ limit for black holes formed through normal single star evolution, but less than the  $10^{5}$ M$_{\odot}$ seen in the smallest galactic nuclei).

The difficulties of stellar dynamics has prompted a search for accretion constraints on the presence of intermediate-mass black holes. As pointed out by Maccarone (2004) and Maccarone et al. (2005), deep radio searches may be a very effective way to detect intermediate-mass black holes in globular clusters and related objects. Indeed, for a given X-ray luminosity, supermassive mass black holes produce far more radio luminosity than stellar-mass black holes. The relation between black hole mass and X-ray and radio luminosity empirically appears to follow a "fundamental plane", in which the ratio of radio to X-ray luminosity increases as the $\sim$0.8 power of the black hole mass (Falcke \& Biermann 1996, 1999; Merloni, Heinz \& Di Matteo 2003; Falcke, Kording \& Markoff 2004). Also, as the luminosity of accretion onto a black hole decreases, the ratio of radio to X-ray power increases \citep{Corbel03,gallo}. 

Accretion theory suggests that the Bondi-Hoyle rate \citep{bondi} overestimates the actual accretion rate by 2-3 orders of magnitude, e.g. \citep{perna}.  Thus, the X-ray luminosities from accretion of the interstellar medium by intermediate-mass black holes in globular clusters are likely to be well below detection limits of current X-ray observatories.  Considering the prediction of \citet{miller}, that the black holes should have about 0.1 per cent of the total cluster mass, the radio luminosities of the brightest cluster central black holes may be detectable with existing instrumentation \citep{mak3}.

Several methods have been considered for proving the existence of these intermediate-mass black holes (in the $10^{2}-10^{4}$ M$_{\odot}$ range), but to date there is no conclusive evidence for their existence.  Searches for radio emission from globular clusters have mostly yielded only upper limits (Maccarone, Fender \& Tzioumis 2005; De Rijcke, Buyle \& Dejonghe 2006; Bash et al. 2008). Although, the cluster G1 in M31 seems to have evidence for harbouring an intermediate-mass black hole \citep{ulve}, including radio detection.

However, beyond globular clusters, there are other possibilities for intermediate-mass black holes.  They may be produced in the core collapses of $\sim 100-1000$-M$_{\odot}$ Population III stars, see e.g. \citep{fryer}. Other good candidates for hosting intermediate-mass black holes are thought to be young dense star clusters (Portegies Zwart \& McMillian 2002; Portegies Zwart et al. 2004; G\"{u}rkan, Freitag \& Rasio 2004) and ultraluminous X-ray sources, whose X-ray luminosities well exceed the Eddington luminosity of a ten solar-mass compact object; for more details see \citep{Kaaret01,zampi}.

The main evidence favouring an intermediate-mass black hole in NGC 6388 is that the observed surface density profile has a power-law shape with a  slope $\alpha = -0.2$ in the inner one arcsecond of the cluster. This slope is shallower than expected for a post core collapse cluster and is consistent with the presence of an intermediate-mass black hole \citep{surf,r}.  The surface density profile provided an estimated mass of 5700 $\pm$ 500 M$_{\odot}$ \citep{Lanzoni07} for the central black hole in NGC 6388 and it motivated us to propose radio observations of the source.

Here, we report on radio observations with the Australia Telescope Compact Array (\textit{ATCA}) of NGC 6388 that led to an upper limit on the mass of the putative
intermediate-mass black hole located at the centre of NGC 6388. In Sec.~2, we describe the analysis of an archival \textit{Chandra} observation and our new  \textit{ATCA} radio observations of NGC 6388. In Sec.~3, we describe  the results of the  X-ray and radio observations. Then we discuss, in Sec.~4, the methodology for setting an upper limit on the mass of a central black hole in NGC 6388 and the uncertainties.

\section{Observations and Analysis}
\subsection{ \textit{Chandra} observation}

NGC 6388 was observed with the Chandra X-Ray Observatory \citep{Weisskopf02} using the Advanced CCD Imaging Spectrometer spectroscopy array (ACIS-S) in imaging mode.  The {\it Chandra} observation (ObsID 5055; PI Haldan Cohn) began on 21 April 2005 02:28:32 UT and had a useful exposure of 45.2~ks.  Although an analysis of the  \textit{Chandra} data was recently published by \citet{nucita}, we improved the localisation of the X-ray sources in the centre of the cluster by removing the pixel randomisation\footnote{http://cxc.harvard.edu/ciao3.4/threads/acispixrand/}. 

The \textit{Chandra} data were subjected to standard data processing (CIAO version
4.1.2 using CALDB version 4.1.4) and then reprocessed to remove pixel randomisation because we are interested in sources in the crowded region near the cluster centre.  We also applied an aspect correction as described on {\it Chandra} aspect webpages\footnote{http://cxc.harvard.edu/ciao/threads/arcsec\_correction/\-index.html\#calc\_corr}.  The total event rate on the S3 chip was less than 1.7~c/s throughout the
observation indicating there were no strong background flares.

\subsection{ \textit{ATCA} observations}

We observed NGC 6388 with the Australia Telescope Compact Array in configuration 6D (baselines up to 6 km) between 24 and 28 December 2008.  The data were obtained simultaneously at 8384 \& 9024 MHz and at 18496 \& 18624~MHz with 16 h and 17 h on-source integration time, respectively. We observed in phase-reference mode; the phase calibrator was 1740-517 and the primary calibrator was PKS 1934-638. The data reduction was performed using the {\sc miriad} software package \citep{miriad} in a standard way.

\section{Results}
\subsection{ \textit{Chandra} Imaging}

We first constructed an image of the full area viewed by the ACIS-S3 chip using all valid events in the 0.3--8~keV band.  The {\it Chandra} aspect uncertainty is $0.6\arcsec$ at 90 per cent confidence\footnote{http://cxc.harvard.edu/cal/ASPECT/celmon}.  To attempt to improve on this we searched for X-ray sources with counterparts \citep{Kaaret02} in the 2MASS catalogue of infrared sources with a J magnitude brighter than 14 \citep{2mass}.  We excluded sources within $40\arcsec$ of the cluster centre due to the source crowding in that region.  We choose X-ray sources which are likely foreground stars by selecting those with soft spectra \citep{Kong07}, specifically those with more counts in the 0.5--1.5~keV band than in the 1.5--6~keV band \citep{Grindlay01}.  There are 10 such sources with a detection significant above $3\sigma$.  Of these 10, one at $\alpha_{\rm J2000} = 17^h 36^m 4.^s55, \delta_{\rm J2000} = -44\degr 45\arcmin 24.\arcsec2$  is located within $0.23\arcsec$ of a 2MASS source and a second at $\alpha_{\rm J2000} = 17^h 36^m 25.^s99, \delta_{\rm J2000} = -44\degr 47\arcmin 53.\arcsec3$ is within $0.37\arcsec$ of a 2MASS source.  There are 636 such 2MASS sources in the search area of $2.25 \times 10^{5}~$\arcsec$^2$.  With 10 X-ray sources, the probability of one chance overlap within $0.4\arcsec$ is 0.0014 and the probability of two chance overlaps is $2 \times 10^{-4}$.  Thus, the {\it Chandra} astrometry appears to be accurate to within $0.4\arcsec$.

\begin{table}
\caption{X-ray sources in the core of NGC 6388 \label{xsources}}
\begin{tabular}{rrclr}
\hline
\, &   S/N &           RA  &           DEC & Counts \\
\hline
1	&	17.7	&	17	36	17.683	&	-44	44	16.78	&	439	\\
2	&	13.9	&	17	36	16.941	&	-44	44	9.90	&	289	\\
3	&	10.3	&	17	36	17.332	&	-44	44	8.36	&	237	\\
4	&	8.6	&	17	36	18.184	&	-44	43	59.53	&	114	\\
5	&	8.1	&	17	36	17.518	&	-44	43	57.17	&	102	\\
6	&	8.1	&	17	36	16.625	&	-44	44	23.43	&	130	\\
7	&	7.1	&	17	36	17.312	&	-44	44	7.10	&	157	\\
8	&	6.9	&	17	36	17.010	&	-44	44	3.06	&	84	\\
9	&	6.8	&	17	36	17.243	&	-44	44	10.38	&	87	\\
10	&	6.1	&	17	36	17.119	&	-44	44	12.83	&	77	\\
11	&	5.3	&	17	36	17.161	&	-44	44	1.92	&	61	\\
12	&	4.5	&	17	36	17.188	&	-44	44	7.61	&	69	\\
13	&	4.4	&	17	36	17.346	&	-44	43	53.73	&	36	\\
14	&	4.2	&	17	36	18.053	&	-44	44	4.30	&	32	\\
15	&	4.1	&	17	36	17.401	&	-44	44	3.19	&	30	\\
16	&	3.7	&	17	36	16.872	&	-44	44	12.85	&	35	\\
17	&	3.6	&	17	36	17.325	&	-44	43	57.21	&	23	\\
18	&	3.4	&	17	36	17.051	&	-44	44	12.44	&	41	\\
\hline
\end{tabular}
\end{table}

We then constructed images of a $38\arcsec \times 38\arcsec$ region centered on the cluster core with pixels which are $0.25\arcsec \times 0.25\arcsec$ in the 0.3--8~keV and 2--10~keV bands.  We searched for sources in the 0.3--8~keV image using the {\it celldetect} tool in {\it CIAO}. A list of detected sources with significance of $3\sigma$ or higher is given in Table~\ref{xsources}.  

Table~\ref{xsources} includes for each source: the source number; S/N -- the significance of the source detection as calculated by {\it celldetect}; RA and DEC -- the position of the source in J2000 coordinates - note that, while the relative positions should be accurate to $0.2\arcsec$, there is an $0.4\arcsec$ overall astrometric uncertainty; Counts - total counts in the 0.3--8~keV band.

\begin{figure}
\resizebox{8.5cm}{!}{
\includegraphics{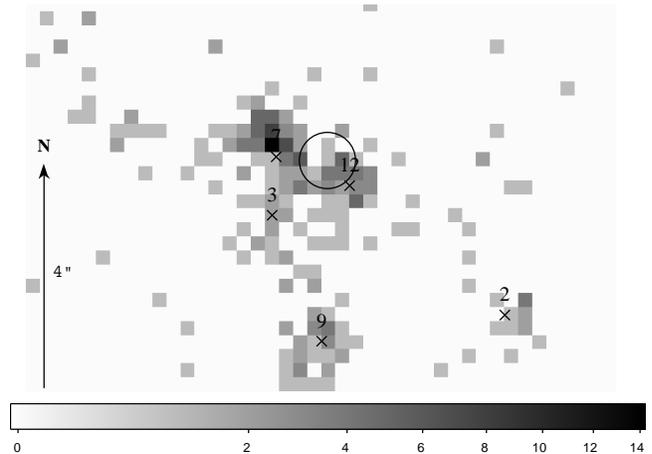}}
\caption{X-ray image of the core of NGC 6388 in the 2-10~keV band with pixel size of $0.25\arcsec$.  The
gray scale indicates X-ray intensity and ranges from 0 to 14 counts per
pixel.  The crosses mark the positions of X-ray sources listed in
Table~\ref{xsources}.  The circle indicates the error circle for the
cluster centre of gravity including the uncertainty relative to the {\it
Chandra} reference frame.}
\label{chandra_image} 
\end{figure}

Fig.~\ref{chandra_image} shows an X-ray image of the core of NGC 6388 in the 2--10~keV band with the position of the cluster center of gravity, $\alpha_{\rm J2000} = 17^h 36^m 17.^s23, \delta_{\rm J2000} = -44\degr 44\arcmin 7.\arcsec1$ \citep{Lanzoni07}, superimposed.  The radius of the circle is $0.5\arcsec$ which is equal to the sum in quadrature of the {\it Chandra} aspect uncertainty of $0.4\arcsec$ (see above) and the uncertainty of $0.3\arcsec$ in the position of the centre of gravity. Two of the {\it Chandra} sources lie near the centre of gravity error circle.  The error circle for source \#12 overlaps the edge.   Source \#7 lies $0.9\arcsec$ from the centre of gravity.  If we use the coordinates calculated from the 2--10~keV image (and not 0.3--8 keV as above), then source \#12 moves inside the  centre of gravity error circle, while source \#7 moves further away.

\subsection{X-ray spectra of source 12 and 7}

We extracted X-ray spectra for these two sources using circular extraction regions with radii of 1.5 pixels centred on the coordinates given in Table~\ref{xsources}.  The extraction radius is smaller than usual for {\it Chandra} sources, but this is necessary given the source crowding.  The 80 per cent encircled power radius is $\sim$0.7 arcsec\footnote{ http://cxc.harvard.edu/proposer/POG/pdf/MPOG.pdf; page 93; figure 6.7} thus, the measured source fluxes were increased by a factor of 1.2 to account for the flux outside the extraction region.  We fitted the X-ray spectra using the {\it Sherpa} spectral fitting package and response matrices calculated using the {\tt mkacisrmf} tool in {\it CIAO}. We used the $\chi^2$-Gehrels statistic to evaluate the goodness of fit due to the low numbers of counts in some spectral bins.

A good fit, $\chi^2/{\rm DoF} = 2.9/8$, was obtained for source \#12 with
an absorbed power-law spectrum with a photon index, $\Gamma$=$1.90 \pm 0.45$ and an equivalent hydrogen absorption column density of $N_{H} = (3.8 \pm 1.8)\times 10^{21} \rm \, cm^{-2}$.  The absorbed flux in the 0.3--8~keV band was $2.4 \times 10^{-14} \rm \, erg \, cm^{-2} \, s^{-1}$ and the unabsorbed flux was $4.0 \times 10^{-14} \rm \, erg \, cm^{-2} \, s^{-1}$ in the 0.3--8~keV band and $2.15 \times
10^{-14} \rm \, erg \, cm^{-2} \, s^{-1}$ in the 2--10~keV band.

An absorbed power-law provided a good fit for the spectrum of source \#7,
$\chi^2/{\rm DoF} = 7.8/16$. The best fit parameters were a photon
index, $\Gamma$=$1.66 \pm 0.27 $ and an equivalent hydrogen absorption column
density of $N_{H} = (3.5 \pm 1.3) \times 10^{21} \rm \, cm^{-2}$.
The lower bound on $N_{H} = 2.2 \times 10^{21} \rm \, cm^{-2}$ was fixed
to the Galactic {\sc HI} column density along the line of sight.  The
absorbed flux in the 0.3--8~keV band was $4.8 \times 10^{-14} \rm \, erg
\, cm^{-2} \, s^{-1}$ and the unabsorbed flux was $6.9 \times 10^{-14}
\rm \, erg \, cm^{-2} \, s^{-1}$ in the 0.3--8~keV band and $4.6 \times
10^{-14} \rm \, erg \, cm^{-2} \, s^{-1}$ in the 2--10~keV band. 

\subsection{Radio non-detection}

\begin{figure}
\resizebox{8.5cm}{!}{\rotatebox{270}{
\includegraphics{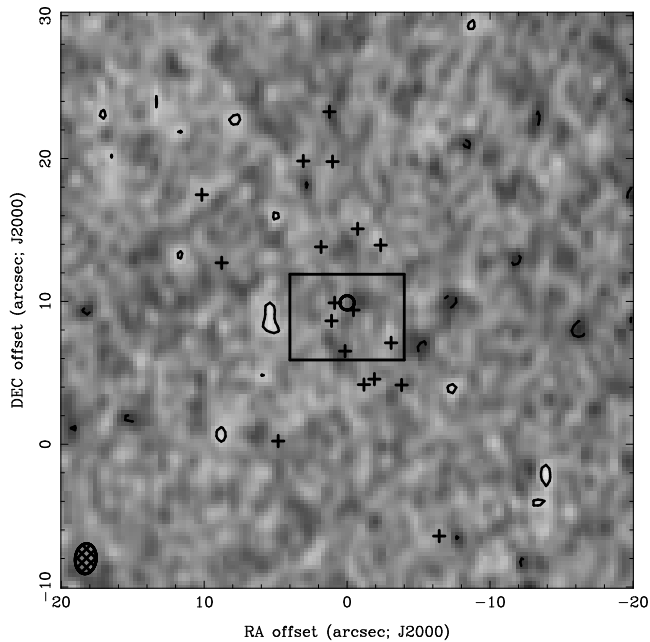}}}
\caption{The naturally weighted  \textit{ATCA} image of NGC 6388 combined at 8.4 GHz \& 9 GHz. The positive contour levels increase by a factor of 2. The first contours are drawn at -81 \& 81 $\mu$Jy/beam. (There are no higher contours than the first ones). The beam size is 2.1 $\times$ 1.5 arcsec at a position angle of 6\degr. The central circle indicates the cluster center of gravity at $\alpha_{\rm J2000} = 17^h 36^m 17.^s23, \delta_{\rm J2000} = -44\degr 44\arcmin 7.\arcsec1$ with a radius  0.5" of the uncertainty. The crosses mark the positions of the  \textit{Chandra} sources. The rectangle indicates the  \textit{Chandra} image showed previously.}
\label{radio_image} 
\end{figure}

After reduction of  the  \textit{ATCA} radio data, we found there are no radio sources detected in association with the cluster centre of gravity nor at the locations of any of the \textit{Chandra} X-ray sources within the cluster (Fig.~\ref{radio_image}). We assumed a flat spectrum (as likely for a low-luminosity accreting black hole) in order to combine the data sets to achieve the best noise level. In naturally weighted maps, by combining 8.4 \& 9 GHz data the rms noise level was 27~$\mu$Jy and combining 18.5 GHz \& 18.6 GHz data the rms was 54~$\mu$Jy. Combining all of the datasets did not lead to better rms as the higher frequency data is more noisy.  Therefore, our best achieved rms noise level at radio frequencies is 27 $\mu$Jy.

\section{Discussion}

Considering the best achieved r.m.s. value obtained from our \textit{ATCA} radio observations and adopting a 3$\sigma$ upper limit, the upper limit on the radio flux of any (undetected) source is estimated to $F_{R} < 81$~$\mu$Jy/beam. The distance of NGC 6388 was adopted from \citet{dales} with the value of 13.2 $\pm$ 1.2~kpc. Therefore, we obtained an upper limit for  the radio luminosity of the putative intermediate mass black hole in NGC 6388 of $L_R < 8.4 \cdot 10^{28}$ erg s$^{-1}$ at 5~GHz ($L_{R}=\nu F_{\nu}$, where we assumed a flat radio spectrum, as is likely for a low-luminosity accreting black hole).

Previously, \citet{nucita} found that three of the \textit{Chandra} sources (\# 12, 7 and 3) coincide with the cluster centre of gravity. They did not attempt to spatially resolve these three sources and therefore obtained one averaged spectra for the three sources. After reanalysing the data and removing the pixel randomisation,  we find that only one source (\#12) is consistent with the location of the centre of gravity (see Section 3.1). In addition,  we successfully measured the spectrum of each source individually. The unabsorbed X-ray flux of source \#12 is $F_{X} = 4.0 \cdot 10^{-14}$ erg cm$^{-2}$ s$^{-1}$, and the corresponding X-ray luminosity is $L_X=8.3 \cdot 10^{32}$ erg s$^{-1}$ in the 0.3--8~keV band. We will use this value in all subsequent analysis.

\subsection{Nature of the innermost X-ray sources}

A study of the colour-colour diagram of the X-ray sources throughout the cluster revealed that most of them seem to be low mass X-ray binary, one source is probably a high-mass X-ray binary and four others are soft sources \citep{nucita}. \citet{nucita} treated the three central \textit{Chandra} sources as one and noted
that the difference between the X-ray flux of these sources measured by \textit{XMM-Newton} in 2003 versus by \textit{Chandra} in 2005 is a factor of only 1.11 . This may indicate that variability of the X-ray sources is small, but there are only two observations. 

From our X-ray data analysis, we managed to localise three separate sources in the centre of the cluster. We studied their X-ray colours and used the source classification method of \citet{jensen}. This revealed that  source \#3 is very soft, probably a cataclysmic variable. Source \#7 seems to be an X-ray binary and source \#12 is either an absorbed source or an X-ray binary. 

The X-ray luminosity of source \#12 is consistent with the accretion luminosity of even a 10 solar-mass black hole in quiescence.  On the basis of the X-ray spectra, there is no real clue to distinguish between an intermediate-mass black hole and low mass X-ray binary. Source \#12 has a power-law spectrum with $\Gamma \approx 1.9$ which is consistent with the spectra of quiescent stellar-mass BH which have $\Gamma \approx 2.0$ (Corbel et al 2006, 2008), but is also consistent with the spectrum expected from a quiescent intermediate-mass black hole.  Thus, the localisation, the source categorisation of \citet{jensen}, and the X-ray spectrum all favour source \#12 as the best candidate to be subjected upon further analysis as a potential intermediate mass black hole.

\subsection{Upper limit on mass of the central black hole}

The fundamental plane of accreting black holes is a relationship between X-ray luminosity, radio luminosity and black hole mass (Merloni et al 2003, Falcke et al. 2004). The relations between mass accretion rate and radio luminosity has been studied for black hole X-ray binaries and low luminosity active galactic nuclei. It is found that the X-ray luminosity is uniquely determined by the black hole mass and radio luminosity via a linear equation in logarithm space. This relationship is referred as "the black hole fundamental plane". It is valid for hard state objects i.e. for objects accreting at low Eddington value. This relation requires radiatively inefficient accretion, like advection dominated or jet dominated accretion flows. 

If one investigates in more detail whether the X-ray source satisfies the criteria for radiative inefficiency, as required for application of the fundamental plane \citep{elmar}, we can conclude, that even if $M=10$~M$_{\odot}$, the ratio of X-ray luminosity to Eddington luminosity,  $L_X / L_{Edd} \simeq 10^{-6}$. So, the source needs to be accreting at a low Eddington-rate. Taking this into account and assuming that the central source is a black hole, we can set constraints on the mass of the black hole using the relationship found by Merloni et al.\ (2003): 

\begin{equation}
\log M=1.28 \log L_R - 0.76 \log L_X - 9.39 
\end{equation}
or the similar relation found by \citet{elmar}:
\begin{equation}
\log M=1.55 \log L_R - 0.98  \log L_X - 9.95. 
\end{equation}  

We note that the relation (Eq. 2.) found by \citet{elmar}, which considered flat spectrum, low-luminosity radio sources, such as those we expect in the centre of dwarf galaxies or globular clusters, more likely reflects the present case. 
However Eq.~1 provides results that are consistent within the uncertainties. 
As Eq.~2 gives a higher mass and has less scatter than Eq.~1 (even than the equation found by \citet{gul}), we use the value given by Eq.~2 in all subsequent analysis. 

Inserting the measured X-ray luminosity ($L_X$) of source \#12 and the measured upper limit on its radio luminosity ($L_R$) into Eq.~2 leads to an upper limit for the mass $M$ of the central black hole of $ < $ 735  $M_{\odot}$. However, we must consider the intrinsic scatter in the measured fundamental plane relation of 0.12 dex ($\sim$ 32 per cent) within one $\sigma$ (and the uncertainty of the distance). Therefore the result is $M < $~735~$\pm$~244~$M_{\odot}$. We adopt the 3-$\sigma$ upper limit, $\approx 1500 M_{\odot}$, as a conservative limit on the black hole mass.

If source \#7 actually lies at the cluster centre of gravity instead source \#12, then we note that source \#7 has a harder spectrum than \#12, being still consistent with a low luminosity accreting black hole and that the difference between their X-ray flux is a factor of 2.  Thus, this would not change the results described above. If none of the X-ray sources are associated with the cluster centre of gravity, then we can not use the fundamental plane to obtain a limit on the black hole mass.

\subsection{Bondi-Hoyle accretion and comparison with G1}

In this section, we investigate more in detail the consistency of the X-ray emission of source \#12 with Bondi-Hoyle accretion and compare NGC 6388 with the cluster G1.
For a black hole of mass $M$ moving with velocity $v$ through gas with hydrogen number density $n$ and sound speed $c_s$, the expected accretion rate via the Bondi-Hoyle process \citep{bondi} is

\begin{equation}
\dot{M} \simeq 4\pi (GM)^{2} m_{p} n(v^2+c_{s}^{2})^{-3/2} 
\end{equation}
where $\rm{m_p}$ is the proton mass \citep{bondieq}.
The expected X-ray luminosity is then
\begin{equation}
L_X \simeq \epsilon\cdot \eta \cdot  \dot{M} c^{2}
\end{equation}
and parametrized for NGC 6388 is
\begin{equation}
L_X \simeq \epsilon \cdot \eta \cdot 8.8 \cdot 10^{36} \left(\frac{M_{\rm{BH}}}{10^3 \rm{M}_{\odot}}\right)^{2} \left(\frac{V}{15 \rm{km/s}}\right)^{-3} \left(\frac{n}{0.1}\right)\frac{\rm{erg}}{\rm{s}}
\end{equation}
where $V$ denotes $(v^2+c_{s}^{2})^{1/2}$, $\epsilon$ is the efficiency for converting accreted mass to radiant energy in the accretion flow of the black hole, and $\eta$ indicates the fraction of the Bondi-Hoyle accretion rate that is accreted by the black hole.

The intracluster gas density can vary in the $\sim 0.1 -1$ cm$^{-3}$ range \citep{bondieq}. For NGC 6388 the gas density is not measured.  Ionised gas with an electron density of $n \simeq 0.1~$cm$^{-3}$ has been detected in the globular cluster 47 Tuc \citep{gasdens}. Recent studies on their stellar populations find 
similar ages for NGC 6388 as 47 Tuc \citep{distance}, therefore we set $n$ = 0.1~cm$^{-3}$ for NGC 6388.  However a higher gas content may be expected for  NGC 6388 \citep{gasngc}.  We also set $v$=0 km/s and $c_{s}$ = 15 km/s. 

Now, the key factors determining the luminosity are the radiative efficiency of accretion ($\epsilon$) and the fraction of Bondi-Hoyle accretion rate that reaches the black hole ($\eta$). Considering our previously obtained 3-$\sigma$ upper limit on mass of $\sim$1500 M$_{\odot}$ and taking $\eta = 1$, we find that the lowest allowed radiative efficiency is $\epsilon \simeq 10^{-4}$, similar values have been found for other systems in the radiatively inefficient regime (Fender et al. 2003, K\"{o}rding et al. 2006b). The highest value of radiative efficiency can be taken as 0.1, i.e. for a radiatively efficient Schwarzschild black hole. So, $\epsilon$ can vary in the $[10^{-4}, 0.1]$ range.

\citet{perna} showed that the lack of detection of isolated neutron stars accreting from interstellar medium implies a fraction of Bondi-accretion of the order of $10^{-3} - 10^{-2}$, see also \citep{pelleg}. However, there is evidence that the central black hole of globular clusters can accrete at a higher fraction of the Bondi-rate than suggested by \citet{perna}.

G1 is an enigmatic star cluster in M 31 thought by some to be a globular cluster, but by others to be the core of a stripped dwarf galaxy \citep{mey} and contains multiple stellar populations. Dynamical evidence in G1 suggested the presence of a 20 000 solar mass black hole (Gebhardt, Rich \& Ho 2002, 2005). X-ray observations revealed an X-ray  source with a 0.2--10 keV luminosity of  $2 \cdot 10^{36}$~erg~s$^{-1}$ \citep{trudo} and \citep{bondieq}. A radio source was also detected at the location of the core of the cluster of G1. The measured radio flux (28~$\mu$Jy) was in good agreement with the predictions (30 and 77 $\mu$Jy) \citep{mak4,ulve}. Although, \citet{Kong09} find that the X-ray emission is still consistent with a single X-ray binary or a collection of X-ray binaries.

\citet{ulve} find the most plausible scenario is that G1 accretes at closer to 0.1 of the Bondi-rate with a radiative efficiency under 0.01. It is therefore possible that the fraction ($\eta$) of the Bondi-Hoyle accretion is as high as 0.1. 

Taking the value - as there is no evidence for pulsar detection in NGC 6388 - obtained by \citet{perna} (see also \citep{pelleg}) as a low end and the value of G1 as a high end, the fraction of the Bondi-rate ($\eta$) can vary on the $[10^{-3}, 0.1]$ range. 

Now, we discuss two scenarios: first, we set the fraction of the Bondi-rate to $\eta= 0.1$ and we assume inefficient accretion ($\epsilon=10^{-4}$) and secondly, we set $\eta=10^{-3}$ and we assume efficient accretion ($\epsilon=0.1$). Using Eq. 5., other cases will lead lower values of black hole mass than derived from the fundamental plane (see Sec. 4.2). In order to reproduce the measured X-ray luminosity (by using Eq. 5.), the first scenario will give an estimate of $\sim$3070~M$_{\odot}$ for the mass of the putative central black hole mass. Such a black hole mass is inconsistent with the derived mass on the fundamental plane. Additionally, this value is even higher than the value of 2600 M$_{\odot}$, which one can obtain considering the raw estimate that, the black hole mass is 1/1000 of the total stellar mass (2.6 $\cdot$ 10$^{6}$~M$_{\odot}$; \citep{Lanzoni07}) of the globular cluster \citep{miller}; therefore this scenario is less favourable. The second scenario will result in a black hole mass of $\sim$970~M$_{\odot}$. This is formally consistent with the one-sigma value derived from the fundamental plane. Although, given the high uncertainties of $\epsilon$ and $\eta$ and the intrinsic scatter of the fundamental plane, our conservative upper limit on the mass of the putative intermediate-mass black hole is $1500$ M$_{\odot}$, the 3-$\sigma$ upper limit derived from the fundamental plane.

\section{Conclusions and Summary}

\citet{Lanzoni07} reported the possible presence of a black hole with a mass of 5700 $\pm$ 500 M$_{\odot}$ at the centre of the globular cluster NGC 6388. \textit{Chandra} and \textit{XMM-Newton} observational data analysis were carried out by \citet{nucita}. Removing the pixel randomisation allowed us to identify a unique source coincident with the cluster centre of gravity with properties consistent with those expected for a black hole accreting at a low rate. With the X-ray detection and optical surface density fit, the only missing piece of the puzzle was a radio detection. On the basis of the X-ray luminosity, \citet{nucita} predicted an upper limit of $<$ 3 mJy radio flux on NGC 6388.  Deep radio observations with the Australia Telescope Compact Array allowed us to reach a sensitivity of 27$\mu$Jy/beam, but did not reveal any radio sources within the cluster.  We interpreted the radio non-detection by using the fundamental plane relating the radio and X-ray properties of black holes accreting at low rates, assuming that the X-ray flux is related to a black hole accretion luminosity. We obtained an upper limit on the black hole mass of $M < $~735~M$_{\odot}\pm$~244~M$_{\odot}$ (1$\sigma$).  Taking into account the uncertainties on the radiative efficiency of accretion and on the fraction of Bondi-Hoyle accretion rate reaching the black hole, we concluded that the centre of NGC 6388 can not host a black hole with a mass in excess of $1500$ M$_{\odot}$ at a 3$\sigma$ confidence level.

\section*{Acknowledgments}

The research leading to these results has received funding from the
European Community's Seventh Framework Programme (FP7/2007-2013) under
grant agreement number ITN 215212 "Black Hole Universe".

\label{lastpage}

\end{document}